\documentclass[11pt]{article}
\usepackage{moriond_losalamos,epsfig,graphicx}
\def\NIMA{{\em Nucl. Instrum. Methods} A}

\def\PLB{{\em Phys. Lett.}  B}

\def\be{\begin{equation}}
\def\ee{\end{equation}}
\def\bea{\begin{eqnarray}}
\def\eea{\end{eqnarray}}

\def\babar{\mbox{\sl B\hspace{-0.4em} {\scriptsize\sl A}\hspace{-0.37em} \sl B\hspace{-0.4em} {\scriptsize\sl A\hspace{-0.02em}R}}}
\begin{document}
\vspace*{4cm}
\title{CHARM MIXING AND LIFETIMES AT BABAR}
\author{A. POMPILI \\ 
(for the \babar\ collaboration)}
\address{Dipartimento di Fisica and I.N.F.N. of Bari,\\ via Amendola 173, I-70126 Bari, Italy}
\maketitle\abstracts{Preliminary limits on the $D^{0}$ mixing parameter $y=\Delta \Gamma /2\Gamma$ are obtained using about $57.8\, \mbox{fb}^{-1}$ of data collected by \babar\ in 2000 and 2001: $y = (1.4 \pm 1.0 \mbox{(stat.)} \mbox{}^{+0.6}_{-0.7}\mbox{(syst.)})\%$. $y$ is extracted, provided that $CP$ is conserved, by measuring separately the $D^{0}$ lifetime for the Cabibbo-suppressed decay modes $K^{-}K^{+}$, $\pi^{-}\pi^{+}$ and the Cabibbo-favored mode $K^{-}\pi^{+}$. Backgrounds are suppressed by $D^*$-tag and particle identification requirements.}
\section{Introduction}

Mixing in the charm sector is a suitable place where effects due to physics beyond the standard model (SM) can show up. The $D^{0}-\overline{D^{0}}$ mixing parameters are defined as $x=\Delta M/\Gamma$ and $y=\Delta \Gamma/2\Gamma$ with $\Delta M\equiv M_{1}-M_{2}$, $\Delta \Gamma\equiv \Gamma_{1}-\Gamma_{2}$ and $\Gamma \equiv \Gamma_{1}+\Gamma_{2}/2$, and where $M_{1,2}$ and $\Gamma_{1,2}$ are the masses and the decay widths for the physical eigenstates $|D_{1,2} \rangle \! = p |D^{0} \rangle \pm q |\overline{D^{0}}\rangle $. In the SM these parameters are predicted to be very small and, even including all the proper corrections, one gets~\cite{sm} $|x|,|y| \! \leq \! 10^{-3}$. However non-SM processes can significantly enhance $x$ whereas final state interactions and SU(3)-breaking can enhance $y$ to a level accessible by the current experimental sensitivity ($10^{-2}$) or in the near future at asymmetric B-factories ($\mbox{few} \times 10^{-3}$).

One possible search for mixing effects can be performed by measuring the lifetime difference between $D^{0}$ decaying, through a Cabibbo-suppressed diagram, to a CP-even eigenstate (such as $K^{-}K^{+}$ and $\pi^{-}\pi^{+}$) and decaying to the Cabibbo-favoured $K^{-}\pi^{+}$ final state~\cite{nir}. Assuming the latter to be an equal mixture of CP-even and CP-odd eigenstates the rate asymmetry for neutral $D$ decays into $CP^{+}$ and $CP^{-}$ eigenstates can be expressed in terms of a lifetime ratio:

\[ 
y_{\, \mbox{\tiny CP}} \equiv \frac{\widehat{\Gamma} (CP^{+})- \widehat{\Gamma} (CP^{-}) }{ \widehat{\Gamma} (CP^{+}) + \widehat{\Gamma} (CP^{-}) } = \frac{ \tau(K^{-}\pi^{+}) }{ \tau(h^{-} h^{+}) } -1 \; , \; h^{-}h^{+}=K^{-}K^{+},\pi^{-}\pi^{+}
\]
where $\tau \! = \! 1/ \widehat{\Gamma}$ and $\widehat{\Gamma}$ is the effective decay rate obtained by fitting to a pure exponential the time-dependent rate for each final state.  The parameter $y_{\, \mbox{\tiny CP}}$ can be expressed~\cite{nir}, by a good ap- proximation, in terms of the CP-violating weak phase $\phi$ and $|q/p|$: $y_{\, \mbox{\tiny CP}} \approx \! y\cos \phi - 1/2 |q/p| x \sin \phi$. Thus, in the CP-conserving limit, $y=y_{\, \mbox{\tiny CP}}$.

Because of the similar topology of the above final states, many systematic uncertainties in the $D^{0}$ lifetimes cancel in the ratio so that the lifetime ratio provides a particularly sensitive measurement of $y$.

\section{Data sample and the \babar\ detector}

The preliminary measurement of $y$ presented here is based on a data sample of $57.8\, \mbox{fb}^{-1}$ collected with the \babar\ detector, at the PEP-II asymmetric $e^{+}e^{-}$ collider, during the years 2000 and 2001. This sample includes data taken both on and off the $\Upsilon(4\mbox{S})$ resonance since the analysis selects events from the $c\overline{c}$ continuum. This analysis uses also Geant4 simulated data, either generic $q\overline{q}$ samples of about $30\, \mbox{fb}^{-1}$ and signal samples of various luminosities.

Detailed description of PEP-II storage ring and the the \babar\ magnetic spectrometer can be found elsewhere~\cite{babar}. The \babar\ subdetectors employed in this analysis are the silicon vertex tracker (SVT), the drift chamber (DCH) and the ring-imaging Cherenkov detector (DIRC) that provide vertexing, tracking and $K/\pi$ separation.

\section{$D^{0}$ candidates selection and proper time reconstruction method}

$D^{0}$ candidates were selected by looking for pairs of charged tracks ($h^{+}h^{-}$) with combined invariant mass close to the $D^{0}$ mass. Each of them was required to satisfy track quality criteria for a good reconstruction. The $\chi^{2}$ probability of the common vertex to which the $h^{+}$, $h^{-}$ were fitted was required to be better than $1\%$. Particle identification criteria were applied to both the tracks to distinguish among the three decay modes and to suppress background. The kaon selection was rather tight and characterized by a pion contamination of less than $3\%$ for momenta less than $3\, \mbox{GeV/c}$. The pion selection included a muon veto. Additional rejection of combinatorial background due to low momenta pions was obtained by cutting on the angle of a pion track in the $D^{0}$ center-of-mass with respect to the $D^{0}$ flight direction.

The $D^{0}$ candidates were required to be produced by $D^{*+} \! \! \rightarrow \! D^{0} \pi^{+}_{s}$ decay. This allowed the introduction of the difference between the $D^{*}$ and the $D^{0}$ reconstructed mass, $\delta m = \! m(h^{+} h^{-} \pi_{s}) - m(h^{+} h^{-})$, which is crucial for background rejection. To increase acceptance, $\pi_{s}$ candidates were required not to contain DCH hits but to contain at least 6 SVT hits. 

\begin{figure}
\centering
\epsfig{figure=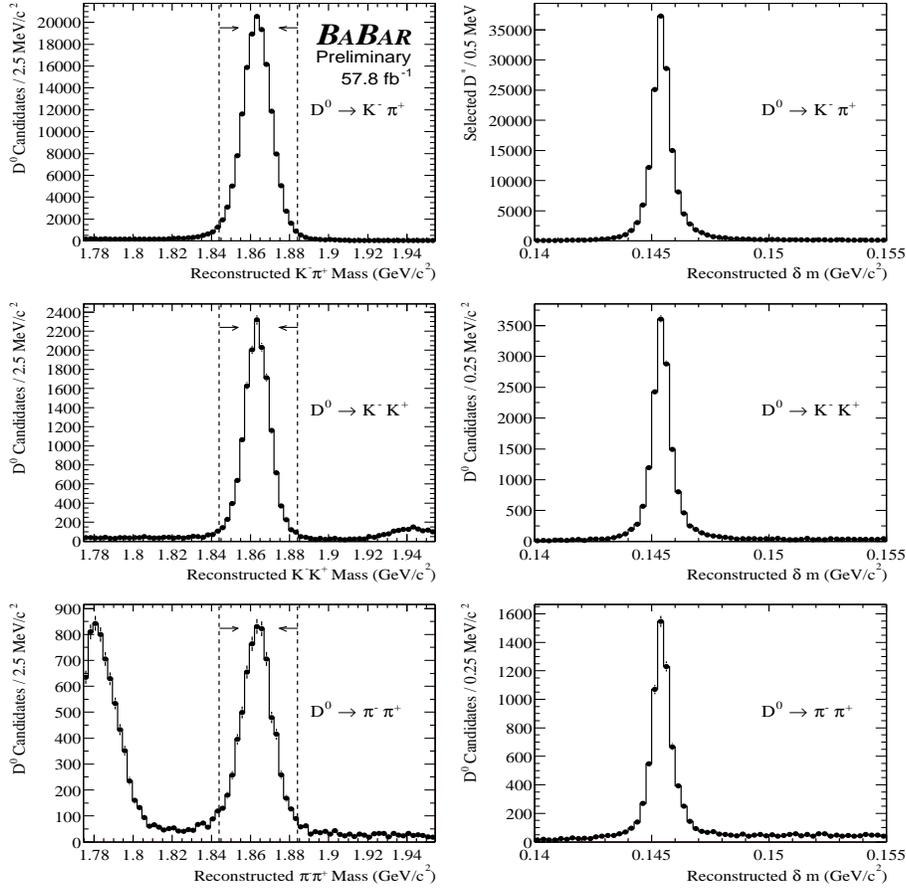,width=12cm,height=12cm}
\caption{On the left the reconstructed $D^{0}$ invariant mass distributions for the three decay modes after the application of the selection criteria. Satellite peaks are due to the reflections of $D^{0} \! \rightarrow \! K^{-}\pi^{+}$ decays deriving from incorrect mass assignment. On the right the corresponding $\delta m$ distributions concern, apart from the $\delta m$-cut itself, the selected $D^{0}$ candidates falling within the mass window indicated by dashed lines into left plots.
\label{fg:mdm}}
\end{figure}

The selection of $D^{*}$ candidates coming from $c\overline{c}$ continuum, namely the rejection of those produced in $B$ meson decays, was obtained by asking for a $D^{0}$ momentum in the $e^{+}e^{-}$  center-of-mass frame greater than $2.5\, \mbox{GeV/c}$. This constrains the $D^{*}$ to be within the interaction region (beam spot) thus allowing a refitting technique with beam spot constraint which highly improves $\delta m$ resolution. The $D^{*}$ decay point was located by pointing the $D^{0}$ momentum vector back to the beam spot and the slow pion, which suffers from multiple scattering, was then refitted to this point and the $D^{*}$ itself obtained by a vertex fit involving the refitted $\pi_{s}$. The $\chi^{2}$ probability of this fit was required to be better than $1\%$. The $\delta m$-cut was performed by opening a $\pm 2 \div \! 3\, \mbox{MeV/}\mbox{c}^{2}$ window around the $\delta m$ distribution (Figure~\ref{fg:mdm}) peak, depending on the quality of the $\pi_{s}$ track. Figure~\ref{fg:mdm} shows the final invariant mass distributions; there were found about $158,000$ $D^{0} \! \rightarrow \! K\pi$, $16,500$ $D^{0} \! \rightarrow \! KK$, and $8,350$ $D^{0} \! \rightarrow \! \pi\pi$ candidates with signal purities of about $99.5\%$, $97.1\%$ and $92.4\%$ respectively, within the $D^{0}$ signal region (indicated by dashed lines in Figure~\ref{fg:mdm}). 

The $D^{0}$ proper time is derived, in the plane transverse to the beams direction, from the flight length defined as the distance between the $D^{0}$ and $D^{*}$ vertices projected on the $D^{0}$ transverse momentum. The choice to measure the flight length in two dimensions derives from the particular beam spot transverse flatness ($\sigma_{y} \approx 5\mu \mbox{m} , \sigma_{x} \approx 120\mu \mbox{m} , \sigma_{z} \approx 9000\mu \mbox{m}$) in connection with the use of the beam spot constraint.

\section{$D^{0}$ lifetime extraction}

An unbinned maximum likelihood fit was used to extract the lifetime for the three $D^{0}$ samples. The likelihood function is divided into two different decay time distributions, one for the signal and one for the background. The former is a pure exponential whereas the latter is the sum of an exponential for the \textit{flying} (charmed) background and a $\delta$ function with a free offset for the \textit{not-flying} (light quark composed) background. Each distribution is smeared by convolution with a proper time resolution function. 

The resolution model, successfully tested for simulated data samples, is a sum of three gaussians in which the first two have a width proportional to the event-by-event proper time error derived from the vertex error matrix and the third, of fixed width, is intended to describe tail contributions. A fourth wide gaussian is added in the smearing of the background function in order to account for long tails. 

To combine the signal and background likelihood functions, the $D^{0}$ reconstructed mass distribution was separately fit to derive the probability for each $D^{0}$ selected candidate to belong to the signal. The sidebands were included in the fit region to better constrain the background.

\begin{figure}
\centering
\begin{tabular}{c@{\hskip -0.24cm}c@{\hskip -0.24cm}c}
\epsfig{figure=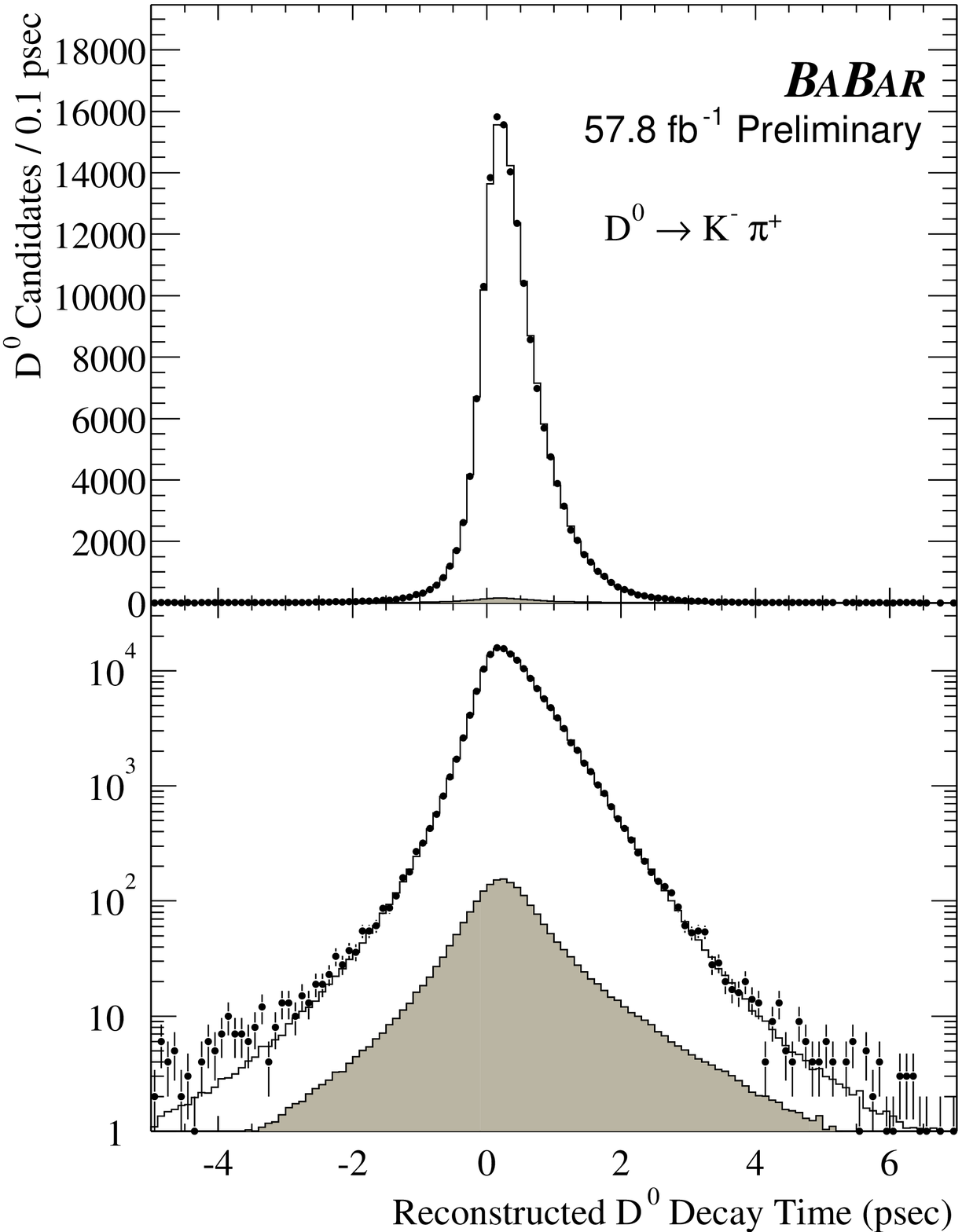,width=5.3cm,height=7.5cm} & \epsfig{figure=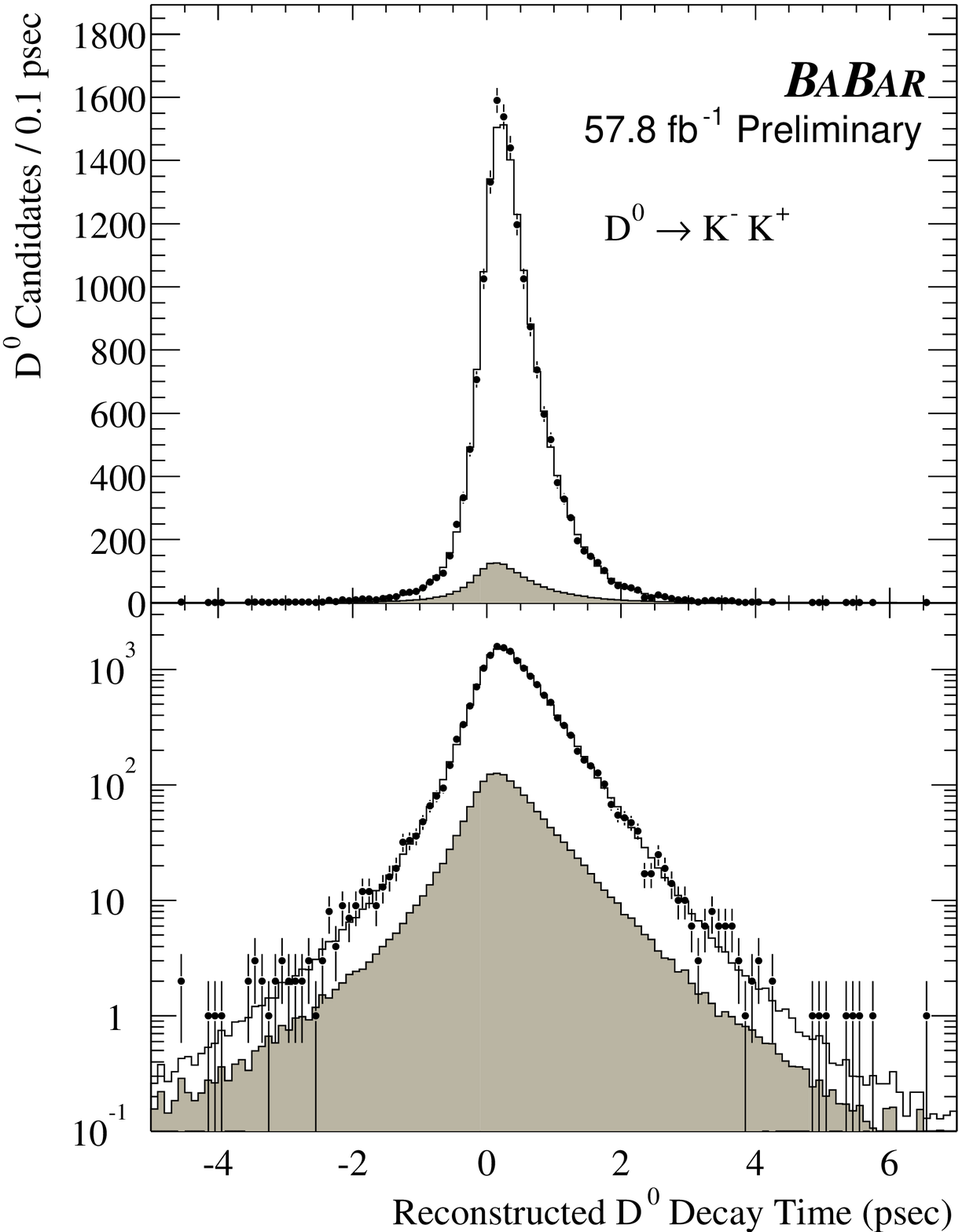,width=5.3cm,height=7.5cm} & \epsfig{figure=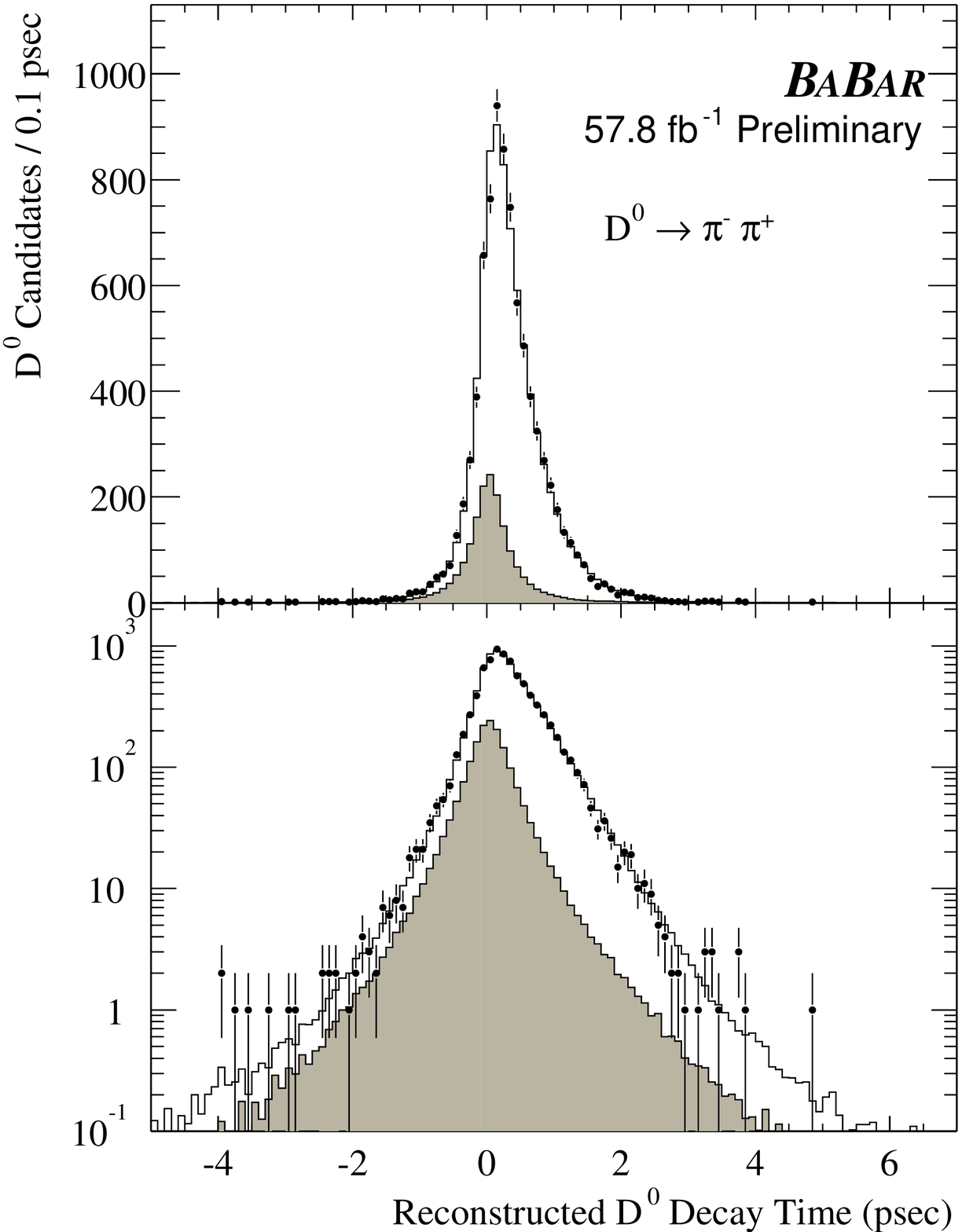,width=5.3cm,height=7.5cm} \\
\end{tabular}
\caption{The white histogram represents the result of the unbinned maximum lilelihood fit described in the text; the gray one is the portion assigned to the background by the fit.
\label{fg:tfits}}
\end{figure}

The results of the lifetime fits are shown in Fugure~\ref{fg:tfits} superimposed on the proper time distributions for the three decay modes.

\section{Lifetime ratio systematics} 

$D^{0}$ lifetime measurement is a high precision measurement since the statistical uncertainty for the Cabibbo-favoured sample is about $1.3~\mbox{fs}$ ($3$ parts per thousand). At this level of precision wide studies of systematic sources (tracking, vertexing, alignment) are needed. 

At this stage we do not present an absolute $D^{0}$ lifetime value because the overall systematic uncertainty (about $3 \div \! 4\, \mbox{fs}$) can be reduced with further studies. We checked the full consistency of the fit result with both our previous check~\cite{hf01} on a sample of $12.8\, \mbox{fb}^{-1}$ and the PDG value~\cite{pdg}, $412\pm 2(stat.)\, \mbox{fs}$ and $412.6\pm 2.8\, \mbox{fs}$ respectively.  

However many systematic effects on lifetime cancel in the lifetime ratio and therefore in $y$.

\scriptsize
\begin{table}[b!]
\begin{center}
\caption{A summary of the systematic uncertainties in the $y$ measurement.}
\vspace{0.4cm}
\renewcommand{\arraystretch}{1.15}
\begin{tabular}{| l | c | c |} 
\hline
$~$ & \multicolumn{2}{|c|}{$y$ Uncertainty (\%)}\\ \hline
Systematic Uncertainty  & $K^-K^+$ &  $\pi^-\pi^+$ \\ \hline 
Tracking                & $ 0.2 $ & $ 0.9 $ \\
Particle Identification & $ 0.2 $ & $ 0.4 $ \\
Background $\&$ $D^{*}$ Fragmentation & $ 0.2 $ & $ 0.6 $ \\
Alignment and Vertexing & \large $ \mbox{}^{ +0.2 }_{ -0.1 } $ & \large $ \mbox{}^{ +0.3 }_{ -0.1 } $ \\ 
Monte Carlo Statistics & \large $ \mbox{}^{ +0.4 }_{ -0.6 } $ & \large $ \mbox{}^{ +0.4 }_{ -0.9 } $ \\ \hline 
Quadrature Sum & \large $ \mbox{}^{ +0.6 }_{ -0.7 } $ & \large $ \mbox{}^{ +1.2 }_{ -1.4 } $ \\ \hline
\end{tabular}
\end{center}
\end{table}
\normalsize

The systematic uncertainties in $y$ were estimated using large signal simulated data samples that were modified by variations in the event selection criteria and suitable variations reflecting the current understanding of the detector, the uncertainties in background level and composition, in beam spot position and size. The $y$ estimation in simulated data shows no bias within statistical errors. Systematic checks of the SVT internal alignment have been performed using $e^{+}e^{-} \! \rightarrow \! \gamma \gamma \! \rightarrow \! 4~prongs$ events which provide a high statistics, zero lifetime control sample. 
All systematic uncertainties in the $y$ parameter are summarized in Table 1.
Major contributions to tracking uncertainties are derived by transverse momentum ($p_{T}$) scale and $\pi_{s}$ resolution uncertainties and, particularly for the $\pi \pi$ mode, by introducing a $q/p_{T}$ costant shift. Particle identification uncertainties are mainly associated to DIRC Cerenkov angle resolution (especially for the $\pi \pi$ mode). Background uncertainties are introduced for the $\pi  \pi$ mode by the level variation of the zero lifetime background and the uncertainty in the portion assigned to the background at lower mass values of the signal region. The major contributions to vertexing/alignment uncertainties are associated to a constant/partially additional smearing in the $y$ position of the interaction point (a variation of beam spot size). However the largest source of uncertainty is from Monte Carlo statistic and will be certainly reduced in the future.

\section{Results and near term prospects}

The measured values of $y$ for the two Cabibbo-suppressed decay modes separately are presented in Table 2 together with their average to be compared with the $y$ limits of similar precision provided by BELLE~\cite{belle} and FOCUS~\cite{focus} collaborations, $(-0.5\pm 1.0 \mbox{}^{+0.7}_{-0.8})\%$ and $(3.42 \pm 1.39 \pm 0.74)\%$ respectively. Our result is consistent with zero but suggests a positive value not incompatible with the FOCUS result. This interesting but by no means conclusive result induces us to perform in the future a new measurement over a rather larger data sample. Indeed the statistical error is determined by the amount of the Cabibbo-suppressed decay events selected.  

\begin{table}
\begin{center}
\caption{A summary of the $y$ results. The first error reported is statistical; the second, systematic.}
\vspace{0.4cm}
\renewcommand{\arraystretch}{1.1}
\begin{tabular}{|l|l@{}c@{}l@{\hskip 2pt}l|} \hline
Decay Mode       & \multicolumn{4}{|c|}{$y$ (\%)} \\ \hline 
$K^- K^+$        & $1.5$ &$\:\pm\:$&$ 1.3 $&$\mbox{}^{+0.6}_{-0.7}$ \\ 
$\pi^- \pi^+$    & $1.0$ &$\:\pm\:$&$ 1.7 $&$\mbox{}^{+1.2}_{-1.4}$ \\ \hline 
average          & $1.4$ &$\:\pm\:$&$ 1.0 $&$\mbox{}^{+0.6}_{-0.7}$ \\ \hline
\end{tabular}
\end{center}
\end{table} 

Limits on $x=\Delta M/\Gamma$ and $y$ derived by the time-dependent analysis of $D^{0} \! \rightarrow \! K^{+}\pi^{-}$ wrong-sign decays are going to be released in the near future by \babar.

\section*{Acknowledgements}

I wish to thank the {\it Rencontres de Moriond} organizing committee for the accommodation grant and the Della Riccia Foundation for its support in 2001.

\end{document}